\documentclass[%
 reprint,
 amsmath,amssymb,
 aps,
]{revtex4-1}

\usepackage{graphicx}
\usepackage{dcolumn}
\usepackage{bm}

\begin{document}

\preprint{APS/123-CTP}

\title{Discrete Boltzmann method with Maxwell-type boundary condition for slip flow}
\author{Yudong Zhang$^{1,2}$}
\author{Aiguo Xu$^{1,3}$}%
 \email{Corresponding author at: Laboratory of Computational Physics, Institute of Applied Physics and Computational Mathematics, Beijing 100088, China. E-mail: Xu_Aiguo@iapcm.ac.cn}

\author{Guangcai Zhang$^{1}$}
\author{Zhihua Chen$^{2}$}

\affiliation{%
 $^{1}$ Laboratory of Computational Physics, Institute of Applied Physics and Computational Mathematics, Beijing, 100088, China\\
 $^{2}$ Key Laboratory of Transient Physics, Nanjing University of Science and Technology, Nanjing 210094, China\\
 $^{3}$ Center for Applied Physics and Technology, MOE Key Center for High Energy Density Physics Simulations, College of Engineering, Peking University, Beijing 100871, China
}%

\date{\today}

\begin{abstract}
The rarefied effect of gas flow in microchannel is significant and cannot be well described by traditional hydrodynamic models. It has been know that discrete Boltzmann model (DBM) has the potential to investigate flows in a relatively wider range of Knudsen number because of its intrinsic kinetic nature inherited from Boltzmann equation.
It is crucial to have a proper kinetic boundary condition for DBM to capture the velocity slip and the flow characteristics in the Knudsen layer.
In this paper, we present a DBM combined with Maxwell-type boundary condition model for slip flow. The tangential momentum accommodation coefficient is introduced to implement a gas-surface interaction model. Both the velocity slip and the Knudsen layer under various Knudsen numbers and accommodation coefficients can be well described. Two kinds of slip flows, including Couette flow and Poiseuille flow, are simulated to verify the model.
To dynamically compare results from different models,
the relation between the definition of Knudsen number in hard sphere model and that in BGK model is clarified.


\begin{description}
\item[PACS numbers]
 51.10.+y, 47.11.-j, 47.45.-n.
\item[Key words]
Rarefied gas, Discrete Boltzmann method, Maxwell-type boundary, slip flow.
\end{description}
\end{abstract}

\keywords{Suggested keywords}
\maketitle


\section{\label{Introduction}Introduction}
In recent years, the development of natural science and engineering technology has moved towards miniaturization. One of the most typical examples is Micro-Electro- Mechanical System (MEMS)\cite{Gad2001Flow,Zhang2005Gas,Bao2005Linear,ChingShen2005Rarefied}. It is extremely important to investigate the underlying physics of unconventional phenomena at the micro-scale. Those unconventional phenomena cannot be explained by the traditional macro-model and has become a key bottleneck limiting the further development of MEMS. Among these unconventional physical problems, the gas flow and heat transfer characteristics at the mico-scale are especially critical.

Due to the reduction of the geometric scale, the mean free path of gas molecules may be comparable to the length scale of the device. The Knudsen number ($Kn$), a dimensionless parameter used to measure the degree of rarefaction of the flow and defined as the ratio of the mean free path of molecules to characteristic length of the device,  may be larger than 0.001 and reaches slip-flow regime(0.001 $<$ $Kn$ $<$ 0.1 ) or even transition-flow regime (0.1 $<$ $Kn$ $<$ 10). As we know, Navier-Stokes (NS) equations are applicable to continuum flow ($Kn$ $<$ 0.001) where the continuum hypothesis is acceptable. The slip boundary models based on kinetic theory should be adopted to describe the slip-flow regime. While in the transition-flow region, the NS equations totally fail to calculate viscous stress and heat flow accurately and higher order equations such as Burnett equation\cite{Burnett1935The,Chenweifang2016} and Grad's 13-moment equations\cite{grad1949kinetic} are needed.

In practical, gas flow in a microchannel may encounter continuum, slip and transition regimes simultaneously. Traditionally macro-scale models cannot apply to such a broad range of Knudsen numbers by only one set of equations. Besides, the numerical solutions of higher order macro-equations are difficult to obtain because of the numerical stability problem \cite{Bao2005Linear}. It is commonly accepted that Direct Simulation Monte Carlo (DSMC)\cite{Bird2003Molecular,ChingShen2005Rarefied} is a accurate method for rarefied gas flow which has been verified by experimental data. However, the computation cost in numerical simulations is too expensive for low speed gas flow. To reduce the huge ratio of the noise to the useful information, extremely large sample size is needed. Although, the information preservation (IP) method was presented to treat this problem\cite{Fan2001Statistical}, the contradiction between the noise problem and sample size has not been well solved.

It has been know that rarefied gas dynamics are represented properly by the Boltzmann equation due to its kinetic nature. That is continuum, slip and transition regimes can be described by one equation. Unfortunately, the Boltzmann equation is a 6-dimensional problem and the computational cost is formidable to solve such an equation by numerical method directly. In order to alleviate the heavy computational burden of directly solving Boltzmann equation, a variety of Boltzmann equation-based methods, such as the unified gas-kinetic scheme (UGKS)\cite{Xu2004Microchannel,Xu2016A}, the discrete velocity method (DVM)\cite{Yang2016Numerical}, the discrete unified gas-kinetic scheme (DUGKS)\cite{Guo2013Discrete,Guo2015Discrete}, the lattice Boltzmann method (LBM)\cite{Zhang2005Gas,Guo2007An,Watari2009,Shan2006Kinetic,Sofonea2005Boundary,Tang2008Lattice,Li2011Lattice,li2016lattice,Ansumali2002Kinetic,Succi2002Mesoscopic,Niu2004A,Tang2005Lattice,Guo2007Discrete}, have been presented and well developed. Recently, Discrete Boltzmann Method (DBM) has also been developed and widely used in various complex flow simulations\cite{Xu2015Progess,Xu2016Progess,Xu2016Complex}, such as multiphase flows\cite{Gan2015discrete}, flow instability\cite{lai2016nonequilibrium,chen2016viscosity}, combustion and detonation\cite{lin2016double,zhang2016kinetic}, etc.
From the viewpoint of numerical algorithm, similar to finite-different LBM, the velocity space is substituted by a limited number of particle velocities in DBM. However, the discrete distribution function in DBM satisfy more moment relations which make it fully compatible with the macroscopic hydrodynamic equations including energy equation. The macroscopic quantities, including density, momentum, and energy are calculated from the same set of discrete distribution functions.
From the viewpoint of physical modeling,
beyond the traditional macroscopic description,
the DBM presents two sets of physical quantities so that the nonequilibrium behaviors can have a more complete and precise description. One set includes the dynamical comparisons of nonconserved kinetic moments of distribution function and those of its corresponding equilibrium distribution function. The other includes the viscous stress and heat flux. The former describes the specific nonequilibrium flow state, the latter describes the influence of current state on system evolution. The study on the former helps understanding the latter and the nonlinear constitutive relations\cite{Xu2014Progess}.
The new observations brought by DBM have been used to understand the mechanisms for formation and effects of shock wave, phase transition, energy transformation and entropy increase in various complex flows\cite{Lin2014Polar,Gan2015discrete,zhang2016kinetic}, to study the influence of compressibility on Rayleigh-Taylor instability\cite{lai2016nonequilibrium,chen2016viscosity}. In a recent study, it is interesting to find that the maximum value point of the thermodynamic nonequilibrium strength can be used to divide the two stages, spinnodal decomposition and domain growth, of liquid-vapor separation.

Some of the new observations brought by DBM, for example, the nonequilibrium fine structures of shock waves, have been confirmed and supplemented by the results of molecular dynamics\cite{Liu2016Molecular,Liu2017Molecular,Liu2017Recent}. It should be pointed out that the molecular dynamics simulations can also gives microscopic view of points to the origin of the slip near boundary, such as the non-isotropic strong molecular evaporation flux from the liquid\cite{Kang2008Thermal}, which might help to develop more physically reasonable mesoscopic models for slip-flow regime.

In order to extend DBM to the micro-fluid, it is critical to develop a physically reasonable kinetic boundary condition. Many efforts have been made to devise mesoscopic boundary condition for LBM to capture the slip phenomenon\cite{Ansumali2002Kinetic,Succi2002Mesoscopic,Niu2004A,Tang2005Lattice,Guo2007Discrete,Zhang2005Gas}. However, the previous works are most suitable for two-dimensional (2D) models with a very small number of particle velocities and can not directly applicable to the DBM. On the other hand, those boundary conditions fail to capture flow characteristics in the Knudsen layer so the effective viscosity or effective relaxation time approach needs to be adopted\cite{Guo2007An,Tang2008Lattice}. Besides, the results of LBM and DBM should be verified by the results of continuous Botlzmann equation. In 2009, Watari\cite{Watari2009} gave a general diffuse reflection boundary for his thermal LB model\cite{Watari2006} and investigated the velocity slip and temperature jump in the slip-flow regime. Then, in his sequent work\cite{Watari2010}, he compared the relationship between accuracy and number of particle velocities in velocity slip. Two types of 2D models, octagon family and D2Q9 model, are used. It was found that D2Q9 model fails to represent a relaxation process in the Knudsen layer and the accuracy of the octagon family is improved with the increase in the number of particle velocities. However, all the boundary conditions were set as diffuse reflection wall and the tangential momentum accommodation coefficient (TMAC) was not taken into account.

Because of the dependence of the mean free path on microscopic details of molecular interaction, especially the collision frequency, the Knudsen number may have different values in various interaction models for the same macroscopic properties.

In this paper, we first clarify the definitions of Knudsen number and the connection between the hard sphere model and BGK model for three-dimensional (3D) condition so that the results obtained from various models can be compared dynamically. Then a general Maxwell-type boundary condition
for DBM is represented and accommodation coefficient is introduced to implement a gas-surface interaction model. Two kinds of gas flows, Couette flow and Poiseuille flow, in a microchannel are simulated. In the section of Couette flow, the relation between the analysis solutions based on hard sphere and BGK model are verified. The simulation results with various Knudsen numbers and accommodation coefficients are compare with analytical ones based on linear Boltzmann equation in the literature not only on the velocity slip but on the Knudsen profiles. While in the section of Poiseuille flow, the simulation results are compare with analytical solution based on Navier-Stokes equation and the second order slip boundary condition.

\section{\label{Methods}Models and Methods}
\subsection{\label{Knudsen number}Definition of Knudsen number}

The Knudsen number is defined as the ratio of the free path of molecules ($\lambda$) to the characteristic length ($L$),
\begin{equation}\label{lambdaHS}
Kn=\frac{\lambda}{L}.
\end{equation}
Throughout the paper, we consider the characteristic length $L$ as unit, so the Knudsen number $Kn$ is equal to the value of $\lambda$.

For the hard sphere collision model, the molecules are considered as hard spheres with diameter $d$, the mean free path of molecules $\lambda_{HS}$ can be calculated by
\begin{equation}\label{lambdaHS}
\lambda_{HS}=\frac{1}{\sqrt{2}n\pi d^2},
\end{equation}
where $n$ is the number density of molecules\cite{ChingShen2005Rarefied}.
According to Chapman and Enskog\cite{Chapman1953The}, the viscosity coefficient $\mu$ of hard sphere molecules can be expressed by
\begin{equation}\label{muHS}
\mu  = \frac{5}{{16}}\frac{{\sqrt {m k T/\pi } }}{{{d^2}}},
\end{equation}
where $k$ is the Boltzmann constant, $m$ is the molecular mass, and $T$ is the temperature. It should be note that gas constant $R$ can be expressed by $R=k/m$.
Combining the state equation of ideal gas ($p=\rho R T$), we have the following relationship between $\lambda_{HS}$ and macroscopic quantities:
\begin{equation}\label{Eq:lambdaHS2}
{\lambda _{HS}} = \frac{4}{5}\frac{\mu }{p}\sqrt {\frac{{8RT}}{\pi }}.
\end{equation}

For the BGK model, the mean free path of molecules $\lambda _{BGK}$ is defined as
\begin{equation}\label{Eq:lambdaBGK}
{\lambda _{BGK}} = \tau \overline{c},
\end{equation}
where $\tau$ is the reciprocal of collision frequency and called relaxation time,  $\overline{c}$ is the average thermal speed \cite{ChingShen2005Rarefied}. The definition of $Kn$ in DBM is in accordance with the definition here.

According to the kinetic theory of gas molecules, $\overline{c}$ is expressed by
\begin{equation}\label{thermalspeed}
  \overline{c} = \sqrt {\frac{{8RT}}{\pi }}
\end{equation}
in 3D case.
From the Chapman-Enskog expansion, we know that $\tau$ has the following relation with macroscopic quantities:
\begin{equation}\label{muBGK}
\mu  = \tau p .
\end{equation}
Consequently, it has
\begin{equation}\label{lambdaBGK2}
  {\lambda _{BGK}} = \frac{\mu}{p}\sqrt {\frac{{8RT}}{\pi }}.
\end{equation}
The comparison of Eq.(\ref{Eq:lambdaHS2}) and Eq.(\ref{lambdaBGK2}) yields the relationship between viscosity-based mean free path $\lambda_{HS}$ and $\lambda_{BGK}$,
\begin{equation}\label{Eq:HSvsBGK}
  {\lambda _{HS}} = \frac{4}{5}{\lambda _{BGK}}.
\end{equation}
\subsection{\label{DBM}Discrete Boltzmann Model}
The 3D discrete Boltzmann model taking into account the effect of the external force was presented based on the thermal model represented by Watari\cite{Watari2006}. The evolution of the discrete distribution function $f_{ki}$ for the velocity particle $\mathbf{v}_{ki}$ is given as
\begin{equation}\label{Eq:fki}
\frac{{\partial {f_{ki}}}}{{\partial t}} + {v_{ki\alpha }}\frac{{\partial {f_{ki}}}}{{\partial {r_\alpha }}} - \frac{{{a_\alpha }({v_{ki\alpha }} - {u_\alpha })}}{T}f_{ki}^{eq} =  - \frac{1}{\tau }({f_{ki}} - f_{ki}^{eq}),
\end{equation}
where the variable $t$ is the time, $r_\alpha$ is the spatial coordinate and $\tau$ is the relaxation-time constant. $a_{\alpha}$ and $u_{\alpha}$ denote the macroscopic acceleration and velocity, respectively, in the $r_{\alpha}$ direction. $T$ denotes the temperature. $f_{ki}^{eq}$ is the local equilibrium distribution function. The subscript $k$ indicates a group of the velocity particles whose speed is $c_k$ and $i$ indicates the direction of the particles. The subscript $\alpha$ indicates an $x$, $y$, or $z$ component.

To recover the NS equations, the local equilibrium distribution function should retain up to the fourth order terms of flow velocity. The discrete local equilibrium distribution $f_{ki}^{eq}$ containing the fourth rank tensor is written as
\begin{eqnarray}
 f_{ki}^{eq} = \rho {F_k}\left[ {(1 - \frac{{{u^2}}}{{2T}} + \frac{{{u^4}}}{{8{T^2}}}) + \frac{1}{T}(1 - \frac{{{u^2}}}{{2T}}){v_{ki\xi }}{u_\xi }} \right.  \nonumber \\
+ \frac{1}{{2{T^2}}}(1 - \frac{{{u^2}}}{{2T}}){v_{ki\xi }}{v_{ki\eta }}{u_\xi }{u_\eta }  {\kern 40pt} \nonumber \\
+ \frac{1}{{6{T^3}}}{v_{ki\xi }}{v_{ki\eta }}{v_{ki\tau }}{u_\xi }{u_\eta }{u_\tau }       {\kern 52pt} \nonumber \\
 \left.{+\frac{1}{{24{T^4}}}{v_{ki\xi }}{v_{ki\eta }}{v_{ki\tau }}{v_{ki\chi }}{u_\xi }{u_\eta }{u_\tau }{u_\chi }} \right],{\kern 8pt}  \label{Eq:fkieq} \end{eqnarray}

The velocity particles $\mathbf{v}_{ki}$ consist of a rest particle and 32 moving particles. Each moving particle has four speeds and can be obtained from the unit vectors in Table \ref{tab:table1} multiplied by difference $c_k$. The speeds $c_k$ of moving particle is determined according to the method presented by Watari  in Ref\cite{Watari2009}.
\begin{table}[b]
\caption{\label{tab:table1}%
 Discrete velocity model, where $\lambda=\frac{1}{\sqrt{3}}$, $\varphi=\frac{1+\sqrt{5}}{2}$, and $\phi=\frac{\sqrt{2}}{\sqrt{5+\sqrt{5}}}$}
\begin{ruledtabular}
\begin{tabular}{ccddd}
$i$ direaction&Unit vector($v_{ix},v_{iy},v_{iz}$) \\
\hline
$i=1-8$&$\lambda(\pm 1,\pm 1,\pm 1)$ \\
$i=9-12$&$\lambda(0,\pm \varphi^{-1},\pm \varphi)$ \\
$i=13-16$&$\lambda(\pm \varphi,0,\pm \varphi^{-1})$ \\
$i=17-20$&$\lambda(\pm \varphi^{-1},\pm \varphi,0)$  \\
$i=21-24$&$\phi(0,\pm \varphi,\pm 1)$ \\
$i=25-28$&$\phi(\pm 1 ,0,\pm \varphi)$ \\
$i=29-32$&$\phi(\pm \varphi,\pm 1 ,0)$ \\
\end{tabular}
\end{ruledtabular}
\end{table}
The $F_k$ in Eq.(\ref{Eq:fkieq}) is the weighting coefficient for the particle velocity $v_{ki}$ and is determined by $c_k$ using the following equations:
\begin{equation}\label{Eq:F0}
{F_0} = 1 - 32({F_1} + {F_2} + {F_3} + {F_4}),
\end{equation}
\begin{eqnarray}
{F_1} = \frac{1}{{c_1^2(c_1^2 - c_2^2)(c_1^2 - c_3^2)(c_1^2 - c_4^2)}} \nonumber {\kern 50pt} \\
\times \left[ \frac{{945}}{{32}}{T^4}
 - \frac{{105}}{{32}}(c_2^2 + c_3^2 + c_4^2){T^3} \right. {\kern 36pt} \nonumber \\
 \left. + \frac{{15}}{{32}}(c_2^2c_3^2 + c_3^2c_4^2 + c_4^2c_2^2){T^2}
- \frac{3}{{32}}c_2^2c_3^2c_4^2T \right],  \label{Eq:F1}
\end{eqnarray}
\begin{eqnarray}
 {F_2} = \frac{1}{{c_2^2(c_2^2 - c_3^2)(c_2^2 - c_4^2)(c_2^2 - c_1^2)}} \nonumber {\kern 50pt} \\
\times \left[ \frac{{945}}{{32}}{T^4}
 - \frac{{105}}{{32}}(c_3^2 + c_4^2 + c_1^2){T^3} \right.   {\kern 36pt} \nonumber \\
 \left.+ \frac{{15}}{{32}}(c_3^2c_4^2 + c_4^2c_1^2 + c_1^2c_3^2){T^2}
 - \frac{3}{{32}}c_3^2c_4^2c_1^2T \right],  \label{Eq:F2}
\end{eqnarray}
\begin{eqnarray}
 {F_3} = \frac{1}{{c_3^2(c_3^2 - c_4^2)(c_3^2 - c_1^2)(c_3^2 - c_2^2)}} \nonumber {\kern 50pt} \\
\times \left[ \frac{{945}}{{32}}{T^4}
- \frac{{105}}{{32}}(c_4^2 + c_1^2 + c_2^2){T^3}  \right.   {\kern 36pt} \nonumber \\
 \left.+ \frac{{15}}{{32}}(c_4^2c_1^2 + c_1^2c_2^2 + c_2^2c_4^2){T^2}
 - \frac{3}{{32}}c_4^2c_1^2c_2^2T \right],    \label{Eq:F3}
\end{eqnarray}
\begin{eqnarray}
{F_4} = \frac{1}{{c_4^2(c_4^2 - c_1^2)(c_4^2 - c_2^2)(c_4^2 - c_3^2)}} \nonumber {\kern 50pt} \\
 \times \left[ \frac{{945}}{{32}}{T^4}
 - \frac{{105}}{{32}}(c_1^2 + c_2^2 + c_3^2){T^3}   \right.   {\kern 36pt} \nonumber \\
 \left.+ \frac{{15}}{{32}}(c_1^2c_2^2 + c_2^2c_3^2 + c_3^2c_1^2){T^2}
 - \frac{3}{{32}}c_1^2c_2^2c_3^2T \right]   .\label{Eq:F4}
\end{eqnarray}

\subsection{\label{Boundary}Boundary condition models}
\begin{figure}
\includegraphics[width=0.5\textwidth]{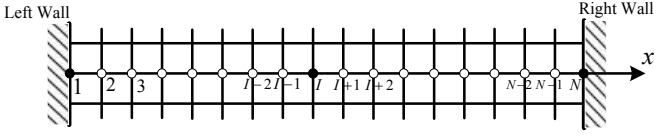}
\caption{\label{Fig1} Schematic of the space grid.}
\end{figure}

\
To solve the evolution equation (\ref{Eq:fki}), finite-difference method is adopted. The spatial derivative is solved by the second-order upwind scheme and time derivative is solved by the first-order forward scheme. Then the evolution equation (\ref{Eq:fki}) can be rewritten as
\begin{eqnarray}
  \nonumber f_{ki}^{t + \Delta t} = f_{ki}^t - {v_{ki\alpha }}\frac{{\partial {f_{ki}}}}{{\partial {r_\alpha }}}\Delta t - \frac{1}{\tau }({f_{ki}} - f_{ki}^{eq})\Delta t \\
  + \frac{{{a_\alpha }({v_{ki\alpha }} - {u_\alpha })}}{T}f_{ki}^{eq}\Delta t  . {\kern 40pt}  \label{Eq:fkiFD}
\end{eqnarray}
The derivation at position $I$ (see Fig.\ref{Fig1})is calculated by
\begin{equation}\label{Eq:2rdFD}
\frac{{\partial {f_{ki}}}}{{\partial {r_\alpha }}} = \left\{ \begin{array}{l}
 \frac{{3{f_{ki,I}} - 4{f_{ki,I - 1}} + {f_{ki,I - 2}}}}{{2\Delta {r_\alpha }}}{\kern 14pt}  {\rm{if}}{\kern 5pt} {v_{ki\alpha }} \ge 0, \\
 \frac{{3{f_{ki,I}} - 4{f_{ki,I + 1}} + {f_{ki,I + 2}}}}{{ - 2\Delta {r_\alpha }}}{\kern 14pt}  {\rm{if}} {\kern 5pt} {v_{ki\alpha }} < 0 .\\
 \end{array} \right.
\end{equation}

For ${v_{ki\alpha }} \ge 0$, the evolution equation (\ref{Eq:fkiFD}) with Eq.(\ref{Eq:2rdFD}) is applied from $I=3$ up to the right wall. However, at the node $I=2$, the second-order upwind scheme in Eq.(\ref{Eq:2rdFD}) is not applicable. The first-order upwind scheme,
\begin{equation}\label{Eq:1stFD}
  \frac{{\partial {f_{ki}}}}{{\partial {r_\alpha }}} = \frac{{{f_{ki,2}} - {f_{ki,1}}}}{{\Delta {r_\alpha }}},
\end{equation}
 is applied there.
For ${v_{ki\alpha }} < 0$, the evolution equation (\ref{Eq:fkiFD}) with Eq.(\ref{Eq:2rdFD}) is applied from the left wall to the node $I=N-2$. Likewise, at the node $I=N-1$, the first-order upwind scheme,
\begin{equation}\label{Eq:1stFD}
  \frac{{\partial {f_{ki}}}}{{\partial {r_\alpha }}} = \frac{{{f_{ki,N - 1}} - {f_{ki,N}}}}{{ - \Delta {r_\alpha }}},
\end{equation}
 is applied.

To solve the value of the distribution function on the left wall for ${v_{ki\alpha }} > 0$  and on the right wall for ${v_{ki\alpha }} < 0$, boundary condition models are required.

\subsubsection{Diffuse reflection boundary}
The complete diffuse reflection model assumes that the molecules leaving the surface with a local equilibrium Maxwellian distribution irrespective of the shape of the distribution of the incident velocity. It can be expressed as
\begin{equation}\label{Eq:rigthdiffuse}
  {f_{ki,N}} = {f_{ki}}^{eq}(\rho _w^R,u_w^R,T_w^R), {\kern 8pt} {v_{ki\alpha }} < 0  ,
\end{equation}
\begin{equation}\label{Eq:leftdiffuse}
  {f_{ki,1}} = {f_{ki}}^{eq}(\rho _w^L,u_w^L,T_w^L), {\kern 8pt} {v_{ki\alpha }} > 0  .
\end{equation}

The equilibrium distribution functions, ${f_{ki}}^{eq}(\rho _w^R,u_w^R,T_w^R)$ and ${f_{ki}}^{eq}(\rho _w^L,u_w^L,T_w^L)$,  are determined from the wall conditions including the velocities and the surface temperatures. Using the zero-mass flow normal to the wall\cite{Watari2009}, the density $\rho _w^R$ and $\rho _w^L$ can be respectively calculated by the following two equations:
\begin{equation}\label{Eq:rhoR}
  \sum\limits_{{c_{ki\alpha }} > 0} {{f_{ki,N}}{c_{ki\alpha }}}  + \rho _w^R\sum\limits_{{c_{ki\alpha }} < 0} {f_{ki}^{eq}(1.0,v_w^R,e_w^R){c_{ki\alpha}}}  = 0  ,
\end{equation}
\begin{equation}\label{Eq:rhoL}
  \sum\limits_{{c_{ki\alpha}} < 0} {{f_{ki,1}}{c_{ki\alpha}}}  + \rho _w^L\sum\limits_{{c_{ki\alpha}} > 0} {f_{ki}^{eq}(1.0,{v_w}^L,{e_w^L}){c_{ki\alpha}}}  = 0  .
\end{equation}

As a result, the distribution function on the left wall (${f_{ki,1}}$) for ${v_{ki\alpha }} > 0$
and on the right wall (${f_{ki,N}}$) for ${v_{ki\alpha }} < 0$ are solved under the diffuse reflection boundary condition.

\subsubsection{Specular reflection boundary}
The specular reflection model assumes that the incident molecules reflect on the wall surface as the elastic spheres reflect on the entirely elastic surface. The component of the relative velocity normal to the surfaces reverses its direction while the components parallel to the surface remain unchanged. As an example, the direction normal to the wall surface parallels to the $x$ axis, then the molecules leave the surface with a distribution as
\begin{equation}\label{Eq:rightSpecular}
  {f_{ki,N}}({v_{kix}},{v_{kiy}},{v_{kiz}}) = {f_{ki,N}}( - {v_{kix}},{v_{kiy}},{v_{kiz}}),  {\kern 4pt} {v_{kix }} < 0,
\end{equation}
\begin{equation}\label{Eq:leftSpecular}
  {f_{ki,1}}({v_{kix}},{v_{kiy}},{v_{kiz}}) = {f_{ki,1}}( - {v_{kix}},{v_{kiy}},{v_{kiz}}),  {\kern 4pt} {v_{kix }} > 0.
\end{equation}

Since the distribution function ${f_{ki,N}}( - {v_{kix}},{v_{kiy}},{v_{kiz}})$ for ${v_{kix}} < 0$  and ${f_{ki,1}}( - {v_{kix}},{v_{kiy}},{v_{kiz}})$ for ${v_{kix}} > 0$ can be solved by Eq.(\ref{Eq:fkiFD}) with Eq.(\ref{Eq:2rdFD}), the distribution function on the right wall (${f_{ki,N}}$) for ${v_{ki\alpha }} < 0$ and on the left wall (${f_{ki,1}}$) for ${v_{ki\alpha }} > 0$ are easy calculated from Eqs.(\ref{Eq:rightSpecular}) and (\ref{Eq:leftSpecular}).

\subsubsection{Maxwell-type boundary}
In practice, the real reflection of molecules on the body surfaces cannot be described properly by complete diffuse reflection or pure specular reflection. So the Maxwell-type reflection model which is composed of the two reflection modes is needed. The TMAC, $\alpha$ is introduced to measure the proportion of complete diffuse reflection\cite{ChingShen2005Rarefied}. The $\alpha$ portion of the incident molecules reflect diffusely and the other ($1-\alpha$) portion reflect specularly. The value of TMAC is used to characterize the degree to which the reflected molecules has adjusted to the tangential momentum of the surface,
\begin{equation}\label{Eq:leftSpecular}
\alpha =\frac{\tau_i-\tau_r}{\tau_i-\tau_w},
\end{equation}
where $\tau_i$ and $\tau_r$ are the tangential components of the momentum fluxes of the incident and reflected molecules, respectively. $\tau_w$ is the tangential momentum fluxes of the molecules in the wall. $\alpha=1$ corresponds to the case of complete tangential momentum accommodation and the molecules reflect with the Maxwellian distribution under wall condtion, $u_w$ and $T_w$. $\alpha=0$ corresponds to the the case when the incident molecules are entirely not adjusted to the conditions of the surface, $\tau_r=\tau_i$.

Under this boundary condition, the distribution function on the right wall, (${f_{ki,N}}$), for ${v_{ki\alpha }} < 0$ and on the left wall, (${f_{ki,1}}$), for ${v_{ki\alpha }} > 0$ are solved by the following equations, respectively,
\begin{eqnarray}
{f_{ki,N}}({v_{kix}},{v_{kiy}},{v_{kiz}}) = \alpha {f^{eq}}(\rho _w^R,u_w^R,T_w^R)    \nonumber {\kern 50pt} \\
+ (1 - \alpha ){f_{ki,N}}( - {v_{kix}},{v_{kiy}},{v_{kiz}}), {\kern 8pt} {v_{ki\alpha }} < 0 , {\kern 20pt}  \label{Eq:rightMaxwell}
\end{eqnarray}
\begin{eqnarray}
{f_{ki,1}}({v_{kix}},{v_{kiy}},{v_{kiz}}) = \alpha {f^{eq}}(\rho _w^L,u_w^L,T_w^L)   \nonumber {\kern 50pt} \\
+ (1 - \alpha ){f_{ki,1}}( - {v_{kix}},{v_{kiy}},{v_{kiz}}), {\kern 8pt} {v_{ki\alpha }} > 0 . {\kern 20pt}  \label{Eq:leftMaxwell}
\end{eqnarray}

\section{\label{Results}Simulation restlts}
\subsection{Couette flow}
Consider a gas flow between two parallel walls, one at $x=-L$  and the other at $x=L$. The two plates are kept at uniform temperature $T_0$ and move with velocity $(0,-v,0)$ and velocity $(0,v,0)$, respectively. Velocity slip becomes more significant with the decrease of the distance between the two plates or with the increase of the mean free path of the molecules, more exactly, with the increase of Knudsen number.

\begin{figure}
\includegraphics[width=0.4\textwidth]{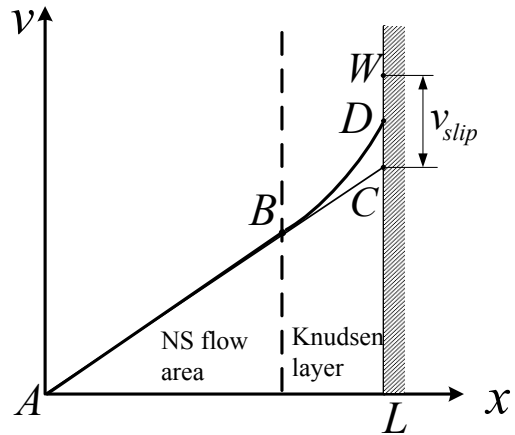}
\caption{\label{Fig2} Schemtic of typical velocity profile for Couette flow with slip effect (right half).}
\end{figure}

The typical velocity profile between parallel plates in the slip-flow regime is depicted in Fig.\ref{Fig2}.
Only right half of the profile is shown because of its antisymmetry. The gas flow away from the wall can be described by NS equations, and the corresponding flow area is referred to as the NS flow area. The flow near the wall possesses pronounced non-equilibrium characteristics,  and the corresponding flow layer is known as the Knudsen layer whose thickness is of the order of the mean free path. In Fig.\ref{Fig2}, the linear portion $A$-$B$, whose gradient is $dv/dx$, corresponds to the NS flow area and the portion $B$-$D$ corresponds to the Knudsen layer. The line $B$-$C$ is extended from the line $A$-$B$ and the point $C$ is the cross point of the extended line with the right wall.
The slip velocity $v_{slip}$ is defined as the difference between the velocity of the right wall (the value the point $W$) and the velocity value of the point $C$. Considering the complete diffuse reflection, Sone gave the relation between $v_{slip}$ and the mean free path $\lambda$ in Ref.\cite{sone2007molecular}.

For the hard sphere model,
\begin{equation}\label{Eq:HSVslip}
  {v_{slip}} = 1.2540\frac{{\sqrt \pi  }}{2}{\lambda_{HS}}\frac{{dv}}{{dx}}.
\end{equation}
For the BGK model,
\begin{equation}\label{Eq:BGKVslip}
  {v_{slip}} = 1.0162\frac{{\sqrt \pi  }}{2}{\lambda_{BGK}} \frac{{dv}}{{dx}}.
\end{equation}
The relationship between $\lambda_{HS}$ and $\lambda_{BGK}$ deduced in Sec.\ref{Knudsen number} is verified by Eq.(\ref{Eq:HSVslip}) and Eq.(\ref{Eq:BGKVslip}) since $1.2540 \lambda_{HS} \approx 1.0162 \lambda_{BGK}$.

Knudsen profile $\Delta v$ is defined as the difference between the curves, $B$-$D$ and $B$-$C$.  Sone\cite{sone2007molecular} gave also the relation between $\Delta v$ and $\lambda$,
\begin{equation}\label{Eq:Knudsenlay}
  \Delta v = {Y_0}(\eta )\frac{{\sqrt \pi  }}{2}{\lambda}\frac{{dv}}{{dx}},
\end{equation}
by introducing the so-called Knudsen layer function, $Y_0(\eta)$,
where $\eta$ is a coordinate transformed from $x$ through the following conversion:
\begin{equation}\label{Eq:eta}
  \eta  = \frac{{x - L}}{{\frac{{\sqrt \pi  }}{2}{\lambda}}}.
\end{equation}

\begin{figure}
\includegraphics[width=0.4\textwidth]{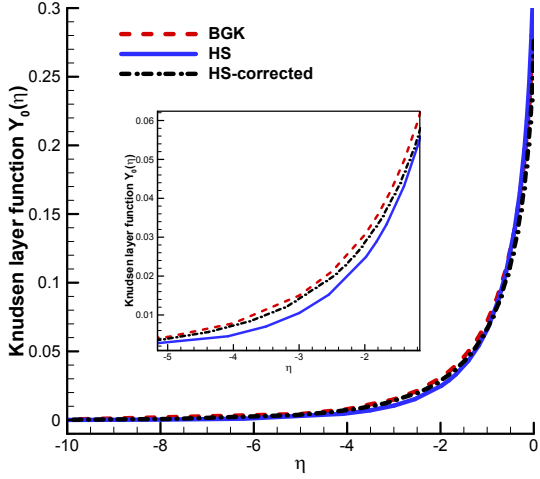}
\caption{\label{Fig3}Profiles of the function $Y_0(\eta)$.
A local enlargement of the curves is shown in the inset.
The solid line is for the Hard Sphere (HS) model. The dashed line is for the BGK, and the dash-dotted line is for the Hard Sphere model with correction (HS-corrected).}
\end{figure}

The Knudsen layer function for hard sphere model, $Y_0^{HS}(\eta)$, and for BGK model, $Y_0^{BGK}(\eta)$, are both shown in Fig.\ref{Fig3}. The correction of the function $Y_0^{HS}(\eta)$ according to the relation in Eq.(\ref{Eq:HSvsBGK}) is also plotted. It can be seen that, the profile of the corrected function is in excellent agreement with the profile $Y_0^{BGK}(\eta)$. As a consequence, the Eq.(\ref{Eq:HSvsBGK}) is revalidated. In addition, the results based on hard sphere model can be compared with those from BGK model under same macro conditions by using the relation of Eq.(\ref{Eq:HSvsBGK}).

Considering the Maxwell-type boundary condition, Onishi\cite{Onishi1974A} gave the expression of slip velocity and Knudsen layer function $Y_0^{(\alpha)}(\eta)$ under various TMACs as
\begin{equation}\label{Eq:velocity_slip2}
  {v_{slip}} = k_s \frac{{\sqrt \pi  }}{2}{\lambda} \frac{{dv}}{{dx}},
\end{equation}
\begin{equation}\label{Eq:Knudsen function2}
  Y_0^{(\alpha )}(\eta ) = \sum\limits_{i = 0}^N {{A_i}{J_i}(\eta )},
\end{equation}
where
\begin{equation}\label{Eq:Jn}
  {J_n}(\eta ) = \int_0^\infty  {{x^n}\exp ( - {x^2} - \frac{\eta }{x})dx},
\end{equation}
$A_i$ and $k_s$ is the coefficient calculated by refined moment methods\cite{sone1973kinetic,Onishi1974A}. According to Onishi\cite{Onishi1974A}, the solutions of $N=7$ are good approximations with high and sufficient accuracy to the exact ones.  The coefficients for partial values of $\alpha$ are listed in Table \ref{tab:table2}. It should be noted that $Y_0^{(\alpha )}(\eta )$ is in complete agreement with ${Y_0}(\eta )$ shown in Fig.\ref{Fig3} when $\alpha=1$ .
\begin{table*}
\caption{\label{tab:table2}Coefficients for several kinds of values of $\alpha$.}
\begin{ruledtabular}
\begin{tabular}{cccccccccc}
 $\alpha$&$A_0$&$A_1$&$A_2$&$A_3$&$A_4$&$A_5$&$A_6$&$A_7$&$k_s$\\ \hline
0.2&-0.8601&2.8585&-10.0275&18.7261&-19.2088&10.4579&-2.8826&0.3036&$8.2248$\\
0.5&-0.6698&2.0540&-7.0263&12.9588&-13.2973&7.2675&-2.0206&0.2141&$2.8612$\\
0.8&-0.5008&1.4022&-4.6499&8.4399&-8.6668&4.7619&-1.3393&0.1431&$1.4877$\\
1.0&-0.3989&1.0437&-3.3750&6.0435&-6.2111&3.4289&-0.9740&0.1047&$1.0162$\\
\end{tabular}
\end{ruledtabular}
\end{table*}

\begin{figure}
\includegraphics[width=0.4\textwidth]{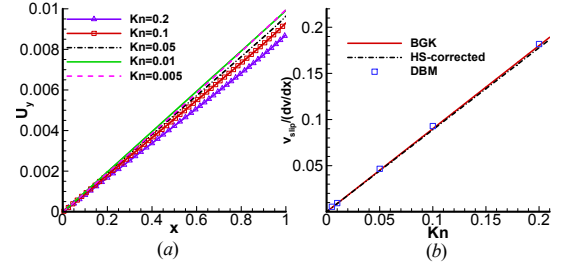}
\caption{\label{Fig4-1} DBM simulation results for different Knudsen numbers under complete diffusion reflection boundary conditions. (a) Velocity profiles for different Knudsen numbers. (b) Slip velocities from DBM and those from by Sone's formulas.}
\end{figure}

The DBM simulation results for the Couette flow with different Knudsen numbers under complete diffusion boundary condition are shown in Fig.\ref{Fig4-1}. Results for different Knudsen numbers are obtained by changing the relaxation-time constant $\tau$ according to Eq.(\ref{Eq:lambdaBGK}).

\begin{figure}
\includegraphics[width=0.4\textwidth]{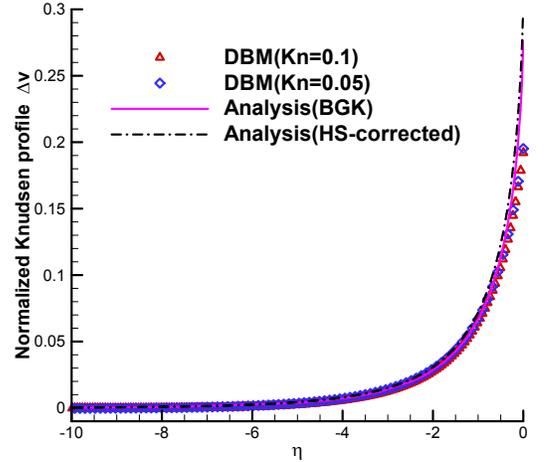}
\caption{\label{Fig4-2}Comparison of normalized Knudsen profiles from DBM simulations and those from analysis.}
\end{figure}

Figure \ref{Fig4-1}(a) shows that the phenomena of velocity slip become more significant with the increase of $Kn$. Comparison of the values of slip velocity normalized by $dv/dx$ between the DBM results and Sone's results is shown in Fig.\ref{Fig4-1}(b).  The two kinds of results  are in excellent agreement with each other. The DBM accurately capture the velocity slip. Besides, the Knudsen layer is also well described by DBM. As shown in Fig.\ref{Fig4-2}, comparison of normalized Knudsen profiles calculated from DBM are also in excellent agreement with Sone¡¯s results. The Knudsen profiles $\Delta v$ in Fig.\ref{Fig4-2} are normalized by $\frac{\sqrt{\pi}}{2} {\lambda}\frac{dv}{dx}$.

\begin{figure}
\includegraphics[width=0.4\textwidth]{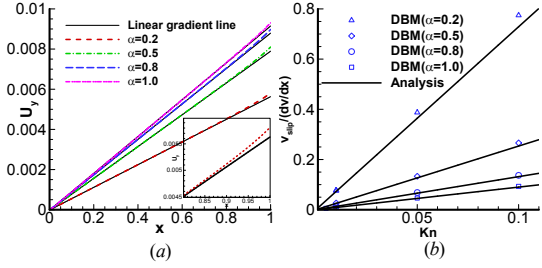}
\caption{\label{Fig4-3} DBM simulation results for different values of $\alpha$ under Maxwell-type reflection boundary conditions.
(a) $u_y (x)$ profiles for different values of $\alpha$. The various dashed lines are for DBM simulation results and the solid lines are for linear fitting results. A local enlargement is shown in the inset.
(b) Comparison of normalized slip velocities from DBM simulations and those from analysis under different values of $\alpha$.}
\end{figure}

\begin{figure}
\includegraphics[width=0.4\textwidth]{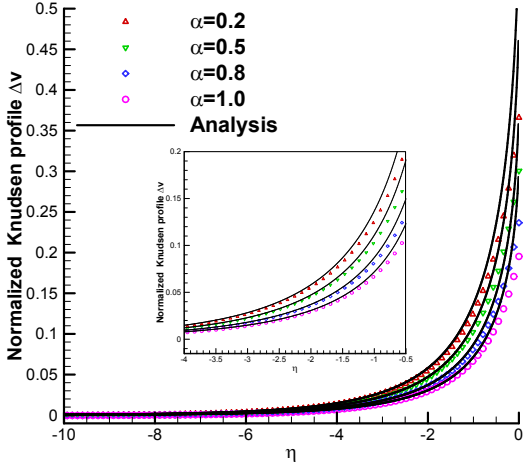}
\caption{\label{Fig4-4} Comparison of normalized Knudsen profiles from DBM simulations with those from analysis for different values of $\alpha$. A local enlargement is shown in the inset.}
\end{figure}

Taking the TMAC ($\alpha$) into consideration, the Maxwell-type boundary condition is adopted in the following simulation. The DBM simulation results with several different values of $\alpha$ are shown in Fig.\ref{Fig4-3}.
From Fig.\ref{Fig4-3}(a), we can see that the phenomena of velocity slip are more significant with the decrease of $\alpha$. The values of velocity slip for different values of $\alpha$ are compared with those given by Eq.(\ref{Eq:velocity_slip2}). Figure \ref{Fig4-3}(b) shows good agreement of DBM simulation results with those of Onishi\cite{Onishi1974A}.
The results of $Y_{0}^{\alpha} (\eta)$  for various values of $\alpha$ are compared in Fig.\ref{Fig4-4}. The DBM results also show good agreement with those of Eq.(\ref{Eq:Knudsen function2}).

\subsection{Poiseuille flow}
Pressure driven gas flow, known as Poiseuille flow,  in a microchannel is also very common in MEMS. In the slip-flow regime, the Navier-Stokes equations with slip boundary condition are applicable. Accurate second-order slip coefficients are of significantly since they directly determines the accuracy of the results given by Navier-Stokes equations\cite{Hadjiconstantinou2003Comment}. The first second-order slip model was presented by Cercignani. Using the BGK approximation he obtained
\begin{eqnarray}
u\mid_{wall}=1.016\theta\frac{\partial u}{\partial y}\mid_{wall}-0.7667\theta ^2\frac{\partial u^2}{\partial y^2}\mid_{wall},
\end{eqnarray}
where $\theta=\mu /p\sqrt{2RT}$. It can be seen, the first-order coefficient is same with Eq.(\ref{Eq:BGKVslip}). Subsequently, Hadjiconstantinou\cite{Hadjiconstantinou2003Comment} improved the model for a hard sphere gas by considering Knudsen layer effects. In his article, the viscosity-based mean free path, $\lambda=\mu /p\sqrt{\frac{\pi RT}{2}}$,
was used. Then he obtained the following slip velocity
\begin{eqnarray}
u\mid_{wall}=1.1466\lambda \frac{\partial u}{\partial y}\mid_{wall}-0.31\lambda ^2\frac{\partial u^2}{\partial y^2}\mid_{wall}.
\end{eqnarray}
However, in our DBM model, viscosity-based mean free path is defined as $\lambda=\mu /p\sqrt{\frac{8RT}{\pi}}$, so the first-order and second-order coefficients should be rescaled by $\pi/4$. The slip velocity formulate should be
\begin{eqnarray}
u\mid_{wall}=0.9004\lambda \frac{\partial u}{\partial y}\mid_{wall}-0.1912\lambda ^2\frac{\partial u^2}{\partial y^2}\mid_{wall}.
\end{eqnarray}
Considering the Maxwell-type boundary, the fully-developed velocity profile can be expressed by
\begin{eqnarray}
u(y) =  - \frac{dp}{dx}\frac{H^2}{2\mu}\left[  - {{(\frac{y}{H})}^2} + \frac{y}{H}  \right.  \nonumber {\kern 55pt} \\
\left. + 0.9004 \frac{2 - \sigma }{\sigma }Kn + 0.3824 Kn^2 \right] . {\kern 10pt}  \label{Eq:Poiseuilleuy}
\end{eqnarray}
where $dp/dx$ is the pressure gradient in the streamwise direction. $H$ is the width of the micro-channel. Nondimensionalize the two sides of Eq.(\ref{Eq:Poiseuilleuy})
by the mean channel velocity $\overline{u}$ gives
\begin{equation}\label{Eq:Poiseuilleuybar}
  U(y) = \frac{{u(y)}}{{\bar u}} = \frac{{ - {{(\frac{y}{H})}^2} + \frac{y}{H} + 0.9004 \frac{{2 - \sigma }}{\sigma }Kn + 0.3824 K{n^2}}}{{\frac{1}{6} + 0.9004 \frac{{2 - \sigma }}{\sigma }Kn + 0.3824 Kn^2}}.
\end{equation}
It is clear that the pressure gradient and the viscosity coefficient vanish.
Firstly, complete diffuse boundary condition is adopted.

\begin{figure}
\includegraphics[width=0.4\textwidth]{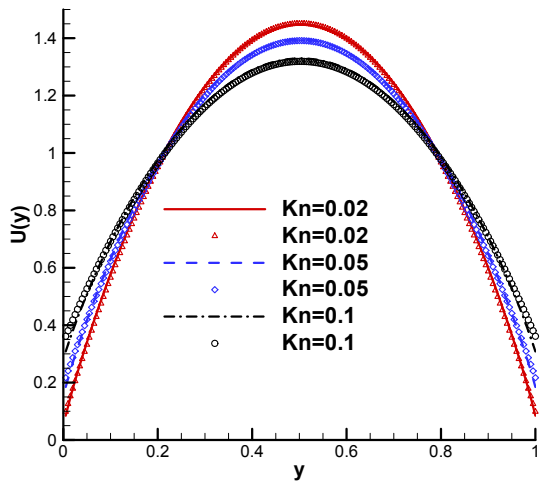}
\caption{\label{Fig5}Comparison of velocity profiles between DBM results and analysis for different Knudsen numbers. Symbols represent the DBM results while lines represent the analytic ones.}
\end{figure}

The simulation results by DBM for different Knudsen numbers are shown in Fig.\ref{Fig5}. It can be found that the velocity slip is significant with the increase of Knudsen number. The nondimensional velocity has a higher maximum value for a smaller Knudsen number. The velocity profile described by Eq.(\ref{Eq:Poiseuilleuybar}) with $\alpha=1$ is also plotted for comparison. The simulation results show good agreement with the expression of Eq.(\ref{Eq:Poiseuilleuybar}).

\begin{figure}
\includegraphics[width=0.4\textwidth]{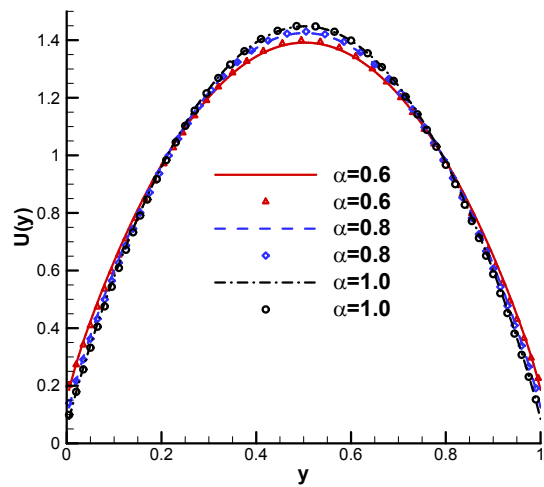}
\caption{\label{Fig6} Comparison of velocity profiles between DBM results and analysis for different values of $\alpha$. Symbols represent the DBM results while lines represent the analytic results.}
\end{figure}

 Considering the behaviours of velocity slip under various TMACs. Then the Maxwell-type boundary condition is used. It can be seen from Fig.\ref{Fig6} that the velocity slip is more significant with the decrease of TMAC and the nondimensional velocity has the highest maximum value when the complete diffuse reflection occurs. It concluded that the effect of Knudsen and TMAC on velocity is in the opposite direction. The numerical results are in well agreement with Eq.(\ref{Eq:Poiseuilleuybar}) for different values of $\alpha$ which verify the accuracy of the Maxwell-type boundary condition.

\section{Conclusion}
A discrete Boltzmann method with Maxwell-type boundary condition for slip flow is presented. The definition of Knudsen number is clarified for DBM. The relation between the Knudsen number based on hard sphere model and that based on BGK model is given. Two kinds of gas flows, including Couette flow and Poiseuille flow, are simulated to verify and validate the new model. The results show that the DBM with Maxwell-type can
reasonably capture both the velocity slip and the flow characteristics in Kundsen layer under  various Knudsen numbers and tangential momentum accommodation coefficients.

\section{acknowledge}
The authors would like to thank Drs. Wei Jiang, Hongwei Zhu, Wei Wang and Ge Zhang for fruitful discussions on discrete Boltzmann modeling of slip flows. We acknowledge support of National Natural Science Foundation of China [under grant nos. 11475028 and 11772064], Science Challenge Project (under Grant No. JCKY2016212A501 and TZ2016002), and Science Foundation of Laboratory of Computational Physics.

\nocite{*}


\end{document}